\documentclass{emulateapj}
\usepackage{graphicx}
\usepackage{natbib}

\newcommand\Eav{{$\left<E\right>$}}

\slugcomment{Accepted by The Astrophysical Journal Letters}

\shorttitle{Is PSR B0656+14 a very nearby RRAT source?}
\shortauthors{Weltevrede et al.}

\begin{document}

\title{Is pulsar B0656+14 a very nearby RRAT source?}

\author{P. Weltevrede\altaffilmark{1}, B.~W. Stappers\altaffilmark{2,1}, J.~M. Rankin\altaffilmark{3,5} \& G.~A.~E. Wright\altaffilmark{4,5}}
\altaffiltext{1}{Astronomical Institute ``Anton Pannekoek'', University of Amsterdam, 
Kruislaan 403, 1098 SJ Amsterdam, The Netherlands (email: {\tt wltvrede@science.uva.nl})}
\altaffiltext{2}{Stichting ASTRON, Postbus 2, 7990 AA Dwingeloo, The Netherlands (email: {\tt stappers@astron.nl})}
\altaffiltext{3}{Physics Department, 405 Cook Physical Science building,
University of Vermont, Burlington, 05405, USA (email: {\tt Joanna.Rankin@uvm.edu})}
\altaffiltext{4}{Astronomy Centre, University of Sussex, Falmer, BN1 9QJ, UK (email: {\tt G.Wright@sussex.ac.uk})}
\altaffiltext{5}{Visiting Astronomer, Astronomical Institute ``Anton Pannekoek'', University of Amsterdam}

\begin{abstract}
The recently discovered RRAT sources are characterized by very bright
radio bursts which, while being periodically related, occur
infrequently. We find bursts with the same characteristics for the
known pulsar B0656+14. These bursts represent pulses from the bright
end of an extended smooth pulse-energy distribution and are shown to
be unlike giant pulses, giant micropulses or the pulses of normal
pulsars. The extreme peak-fluxes of the brightest of these pulses
indicates that PSR B0656+14, were it not so near, could only have been
discovered as an RRAT source. Longer observations of the RRATs may
reveal that they, like PSR B0656+14, emit weaker emission in addition
to the bursts.
\end{abstract}

\keywords{stars: pulsars --- stars: individual (PSR B0656+14) --- radiation mechanism: non-thermal}

\section{Introduction}

\object{PSR B0656+14} is one of three nearby pulsars in the middle-age range in
which pulsed high-energy emission has been detected. These are
commonly known as ``The Three Musketeers'' \citep{bt97}, the other two
being Geminga and PSR B1055--52. PSR B0656+14 was included in a recent
extensive survey of subpulse modulation in pulsars in the northern sky
at the Westerbork Synthesis Radio Telescope (WSRT) by
\citet{wes06}. In the single pulses analysed for this purpose, the
unusual nature of this pulsar's emission was very evident, especially
the brief, yet exceptionally powerful bursts of radio emission.

These extreme bursts of radio emission of PSR B0656+14 are similar to
those detected in the recently discovered population of bursting
neutron stars. These Rotating RAdio Transients (RRATs;
\citealt{mll+06}) typically emit detectable radio emission for less
than one second per day, causing standard periodicity searches to fail
in detecting the rotation period. From the greatest common divisor of
the time between bursts, a period has been found for ten out of the
eleven sources. The periods (between 0.4 and 7 s) suggest these
sources may be related to the radio-quiet X-ray populations of neutron
stars, such as magnetars \citep{wt06} and isolated neutron stars
\citep{hab04}. However, \citet{ptp06} have shown that the estimated
formation rate of magnetars is too low. Furthermore the spectrum of
the only RRAT for which an X-ray counterpart has so far been detected
\citep{rbg+06} seems to be too cool, too thermal and too dim for a
magnetar, but is consistent with a cooling middle-aged neutron star
like PSR B0656+14
\citep{szk+06}. Also the pulse period and the slowdown-rate of PSR
B0656+14, as well as the derived surface magnetic field strength and
characteristic age, are within the range of measured values for RRATs.

\section{Observations}
\label{SctObs}

The results in this paper are based on an archival and a new
observation made using the 305-meter Arecibo telescope on 20 July 2003
and 30 April 2005 respectively. Both observations had a
centre-frequency of 327 MHz and a bandwidth of 25 MHz. Almost 25,000
and 17,000 pulses with a sampling time of 0.5125 and 0.650 ms were
recorded using the Wideband Arecibo Pulsar Processor
(WAPP\footnote{{\tt http://www.naic.edu/{\tiny{$\!\sim$}}wapp}}) for
the 2003 and 2005 observation respectively. The Stokes parameters have
been corrected off-line for dispersion, Faraday rotation and various
instrumental polarization effects.

The data were in some instances affected by Radio Frequency
Interference, but this could relatively easily be removed by excluding
the pulses with the highest root-mean-square (RMS) of the off-pulse noise
(about 1\% of the data in both observations) from further
analysis. The results derived from both observations (of which the one
from 2005 is relatively clean) are very similar, making us confident
in the results. 

The observations are not flux calibrated, but are sufficiently long to
get a pulse profile with high precision. From its shape it follows
that the peak-flux of the profile is about 17 times that of the
integrated flux-density over the entire pulse phase. The average flux
at our observing frequency is estimated to be 7.2 mJy, based on the
measurement of the spectral index ($-0.5$) and flux-density by
\citet{lylg95} at 408 MHz. Therefore the peak-flux of the profile is
approximately 0.12 Jy. The scintillation bandwidth of PSR B0656+14 is
much smaller than the observing bandwidth, so no intensity
fluctuations appear in the data as a function of time due to
interstellar scintillation.

\section{The radio bursts of PSR B0656+14}

\begin{figure}
\epsscale{.99}
\plotone{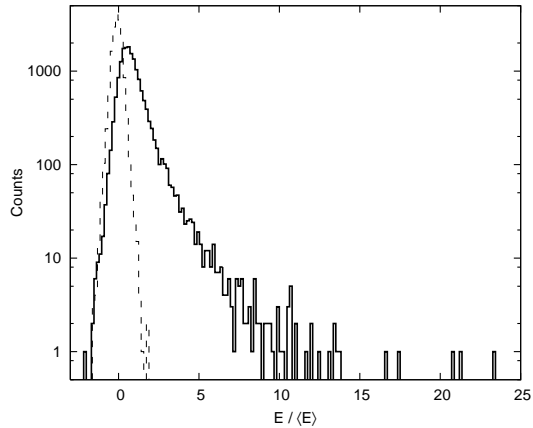}
\caption{\label{Enhists}The pulse-energy distribution of
the 2005 observation of PSR B0656+14 (solid line) and the off-pulse
distribution (dashed line). The brightest pulse is about 116 times
stronger than the average, which is outside the plotted energy-range.}
\end{figure}

To characterize the bright pulses of PSR B0656+14, the pulse-energy
distribution is calculated (see Fig. \ref{Enhists}). In this plot the
energies are normalized to the average pulse-energy {\Eav}. The
brightest measured pulse is 116 {\Eav}, which is exceptional for
regular radio pulsars. Based on the energy of these pulses alone, PSR
B0656+14 would fit into the class of pulsars that emit so-called giant
pulses (e.g. \citealt{cai04}). About 0.3\% of the pulses of PSR
B0656+14 are brighter than 10 {\Eav}, which is the working definition
of giant pulses. Nevertheless, there are important differences between
giant pulses and the bright bursts of PSR B0656+14. The bursts of PSR
B0656+14 have timescales that are much longer than the nano-second
timescale observed for giant pulses (e.g. \citealt{spb+04,hkwe03}), do
not show a power-law energy-distribution, are not confined to a narrow
pulse window and are not associated with an X-ray component. This
suggests differing emission mechanisms for the classical giant pulses
and the bursts of PSR B0656+14. Also the possible correlation between
emission of giant pulses and high magnetic field strengths at the
light cylinder (around $10^5$ Gauss;
\citealt{cst+96}) clearly fails for PSR B0656+14 (766 Gauss). 
However, giant pulses have been claimed in other (slow) pulsars that
also easily fail this test (e.g. \citealt{ke04,kkg+03}) and for
millisecond pulsars a high magnetic field strengths at the light
cylinder seems to be a poor indicator of the rate of emission of giant
pulses \citep{kbmo+06}.

\begin{figure}
\epsscale{.80}
\plotone{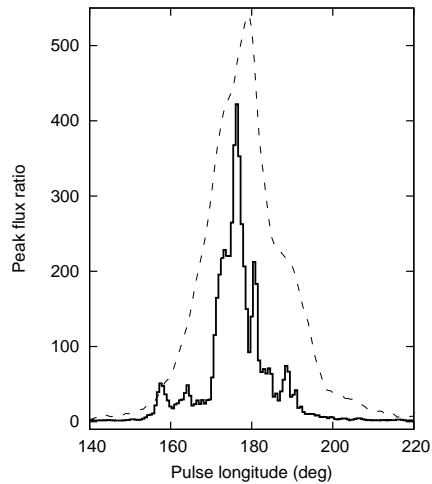}
\caption{\label{lrced}
The dashed line is the average profile of our 2005 observation. The
solid line shows the ratio between the peak-flux of the brightest
burst at each longitude and the average peak-flux of the profile.
}
\end{figure}

The bursts of PSR B0656+14 are even more extreme when we consider
their peak-fluxes (see Fig. \ref{lrced}). The highest measured
peak-flux of a burst is 420 times the average peak-flux of the pulsed
emission, which is an order of magnitude brighter than the giant
micropulses observed for the Vela pulsar \citep{jvkb01} and PSR
B1706--44 \citep{jr02}. Giant micropulses are not necessarily extreme
in the sense of having a large integrated energy (as required for
giant pulses), but their peak-flux densities are very large. Not only
are the bursts of PSR B0656+14 much brighter (both in peak-flux and
integrated energy) than those found for giant micropulses, they are
also not confined in pulse longitude and they do not show a power-law
energy-distribution as the giant pulses and micropulses do.

\begin{figure*}
%\epsscale{.25}
%\plotone{f3a.eps}
%\epsscale{.39}
%\plotone{f3b.ps}
%\epsscale{.35}
%\plotone{f3c.eps}
\begin{center}
\resizebox{!}{0.29\hsize}{\includegraphics[angle=0,trim=0 -15 0 0,clip=true]{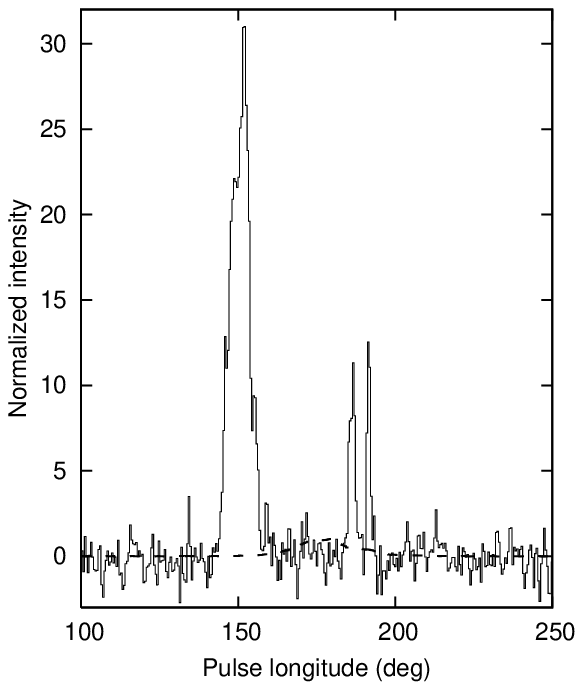}}
\resizebox{!}{0.29\hsize}{\includegraphics[angle=0,trim=0 -30 0 0,clip=true]{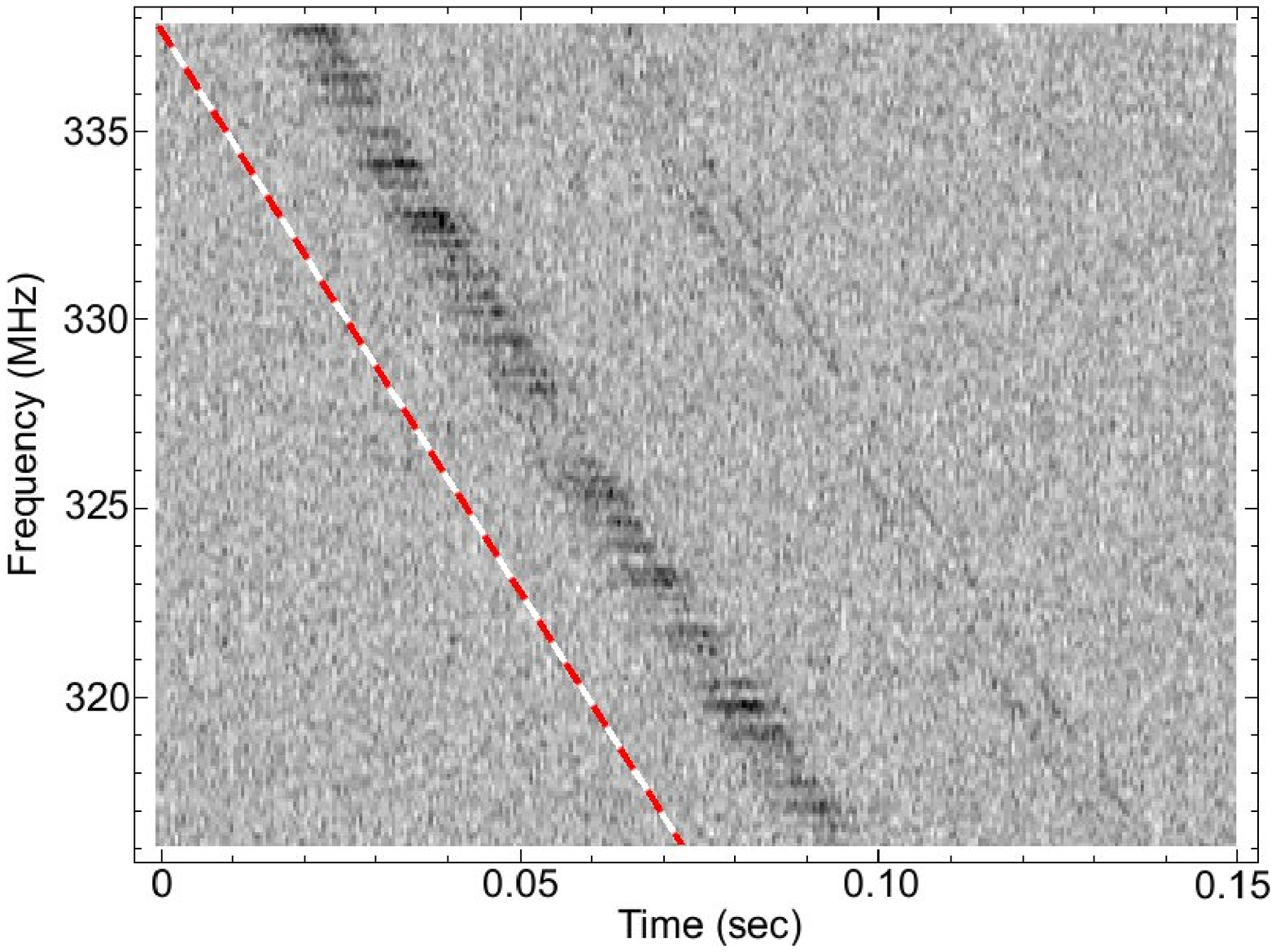}}
\resizebox{!}{0.3\hsize}{\includegraphics[angle=0,trim=0 0 0 0,clip=true]{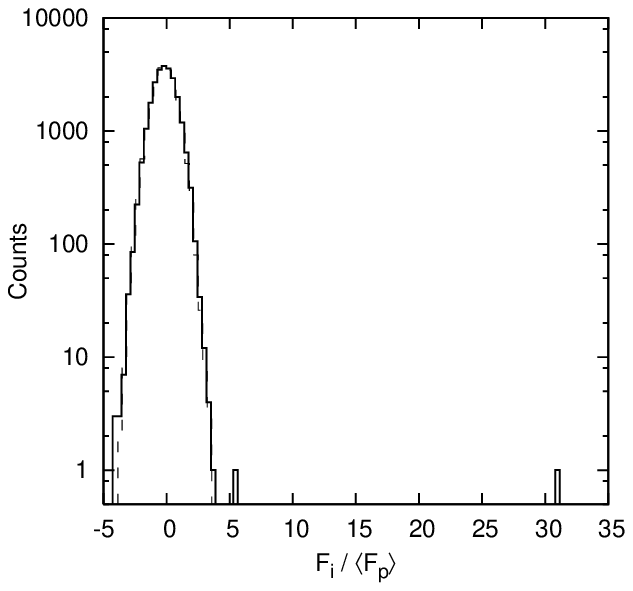}} \vspace*{-5mm}
\end{center}
\caption{\label{megapulse}The bright radio burst detected at the
leading edge of the pulse profile in the 2003 observation. {\bf Left:}
The burst (solid line) compared with the average pulse profile (dashed
line). {\bf Middle:} The same burst, but now with frequency
resolution. The data in this plot is not de-dispersed and its
dispersion track matches exactly what is expected for the known
dispersion measure (DM) of this pulsar (dashed line). {\bf Right: }
The longitude-resolved energy-distribution at the longitude of the
peak of the strong pulse (solid line) and the off-pulse distribution
(dashed line). The peak-fluxes ($F_i$) are normalized to the average
peak-flux of the profile ($\left<F_p\right>$).}
\end{figure*}

At the leading edge of the profile we detected a burst with an
integrated pulse-energy of $12.5\left<E\right>$. What makes this pulse
so special is that it has a peak-flux that is 2000 times that of the
average emission at that pulse longitude (left panel of
Fig. \ref{megapulse}). Its dispersion track exactly matches what is
expected for this pulsar (middle panel of Fig. \ref{megapulse}),
proving that this radio burst is produced by the pulsar. Notice that
the effect of interstellar scintillation is also clearly visible
(different frequency channels have different intensities) and that the
dispersion track is the same for the two pulses in the centre of the
profile. This burst demonstrates that the emission mechanism operating
in this pulsar is capable of producing intense sporadic bursts of
radio emission even at early phases of the profile. There are only two
bursts with a peak-flux above the noise level detected at the
longitude of the peak of this pulse out of the total of almost 25,000
pulses (see right panel of Fig. \ref{megapulse}). This implies either
that these two bursts belong to an extremely long tail of the
distribution, or that there is no emission at this longitude other
than such sporadic bursts.

\section{The RRAT connection}

It is unclear how the extreme pulses of PSR B0656+14 fit into the zoo
of apparently different emission types of radio pulsars. They are
brighter than the giant micropulses, and not constrained to a
particular pulse longitude and, despite being energetic enough, they
are too broad to be characterized as classical giant pulses. However,
the observational facts are that PSR B0656+14 occasionally emits
extremely bright bursts of radio emission which are short in
duration. One cannot help but see the similarities with the RRATs.

\begin{table}
\caption{\label{LuminositiesTable}Comparison of the peak-flux of
the brightest bursts of PSR B0656+14 and those of the RRATs. Here
$S_\mathrm{peak}$ is the peak-flux of the brightest detected burst for
each source, $D$ the distance and $L_\mathrm{peak}=S_\mathrm{peak}d^2$
the peak-luminosity of the brightest detected burst. }
\begin{center}
\begin{tabular}{lrcc}
\hline
\hline
Name & $S_\mathrm{peak}$ & $D$  & $L_\mathrm{peak}$   \\
 & mJy & kpc & Jy kpc$^2$  \\
\hline
B0656+14    & 59000 & 0.288 & 4.1\\
J0848--43   &   100 & 5.5   & 3.0\\
J1317--5759 &  1100 & 3.2   & 11\\
J1443--60   &   280 & 5.5   & 8.5\\
J1754--30   &   160 & 2.2   & 0.77\\
J1819--1458 &  3600 & 3.6   & 47\\
J1826--14   &   600 & 3.3   & 6.5\\
J1839--01   &   100 & 6.5   & 4.2\\
J1846--02   &   250 & 5.2   & 6.8\\
J1848--12   &   450 & 2.4   & 2.6\\
J1911+00    &   250 & 3.3   & 2.7\\
J1913+1333  &   650 & 5.7   & 21\\
\hline  
\end{tabular}
\end{center}
\end{table}

One important question is whether the luminosities of the bursts of
the relatively nearby PSR B0656+14 (288 pc; \citealt{btg+03}) and
those of the RRATs are comparable (there is no indication that the
spatial distributions of RRATs and PSRs are different;
\citealt{mll+06}). The brightest burst we found in the centre of the
profile has a peak-flux that is 420 times the average peak-flux. With
an average peak-flux of 0.12 Jy (see Sect. \ref{SctObs}), this
corresponds to a peak flux of 50 Jy. If one compares luminosities
(Table \ref{LuminositiesTable}), one can see that the brightest burst
of PSR B0656+14 is as luminous as those of four of the eleven RRATs,
and therefore very typical for these sources.

It is interesting to compare not only the luminosities of the bursts,
but also their peak-flux distributions. Although the slope of the top
end of the distribution of PSR B0656+14 is in the range of the giant
pulses (between $-2$ and $-3$), it is better described by a lognormal
than by a power-law distribution. This again suggests that the bright
bursts of PSR B0656+14 are different from the classical giant
pulses. The top end of the RRAT distribution with the highest number
of detections seems to be harder (with a slope $-1$), but for the
other RRATs this is as yet unclear. For instance, the tail of the
distribution of PSR B0656+14 seems to be consistent with the
distribution of the RRAT with the second highest number of detections.

Normal periodicity searches failed to detect the RRATs, which places
an upper limit on the average peak-flux density of weak pulses among
the detected bursts of about 1:200 \citep{mll+06}. Because the
brightest burst of PSR B0656+14 exceeds the underlying peak-flux by a
much greater factor, PSR B0656+14 could have been identified as an
RRAT, were it not so nearby. Were it located twelve times farther away
(thus farther than five of the RRATs), we estimate that only one burst
per hour would be detectable (the RRATs have burst rates ranging from
one burst every 4 minutes to one every 3 hours). The typical burst
duration (about 5 ms) of PSR B0656+14 also matches that of the RRATs
(between 2 and 30 ms).

\begin{figure*}
\epsscale{0.55}
\plotone{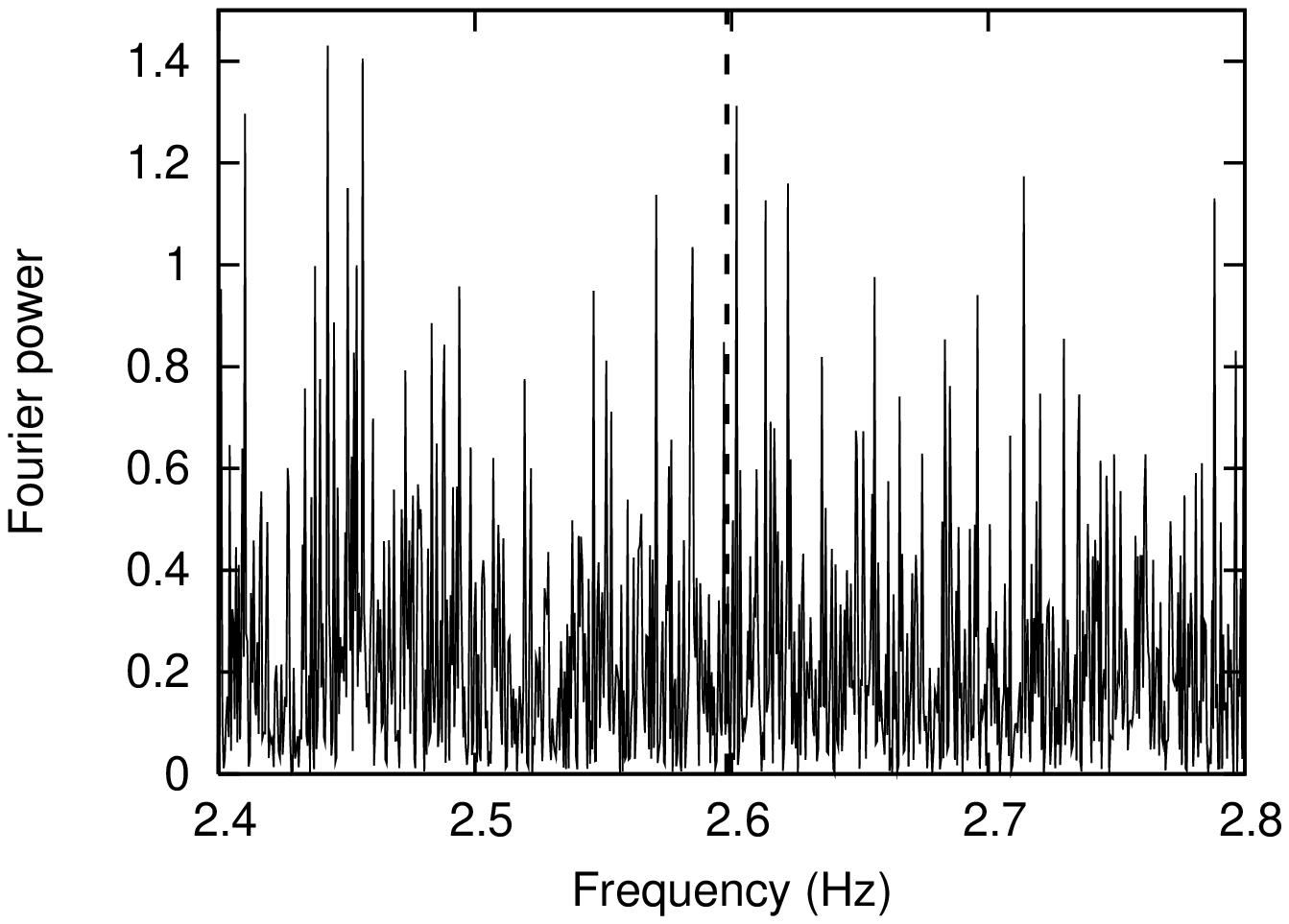}
\epsscale{0.27}
\plotone{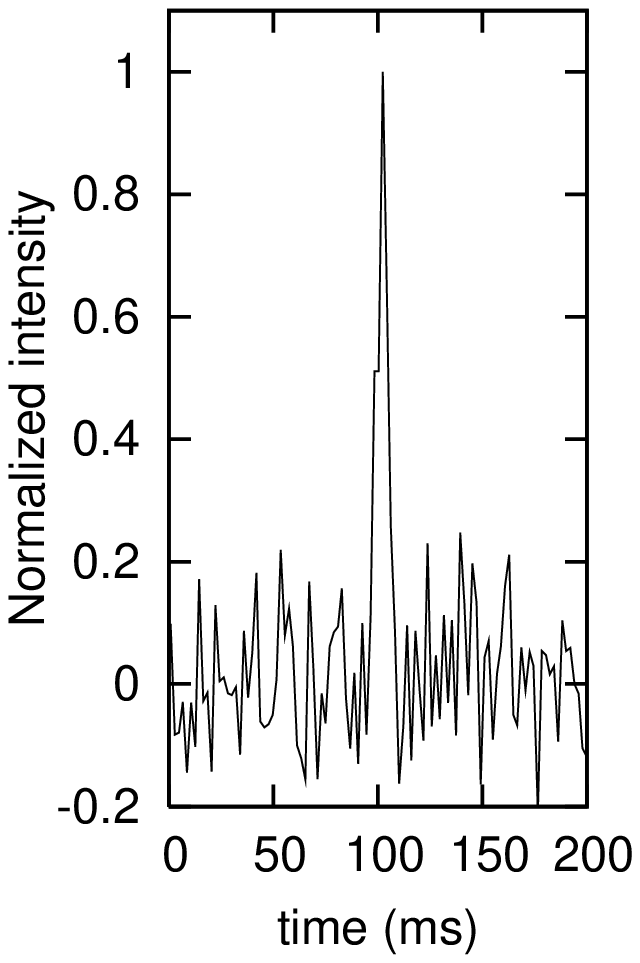}
\caption{\label{distribution}{\bf Left:} The power-spectrum of 
the first 35 minutes of the 2005 observation of PSR B0656+14 with
artificially increased noise corresponding to a twelve times larger
distance of this pulsar. The dashed lines indicates the rotation
frequency of this pulsar. {\bf Right:} The brightest recorded pulse in
the same piece of data, again for a twelve times larger distance.}
\end{figure*}

Were PSR B0656+14 twelve times more distant, the RMS of the noise
would increase by a factor 144 relative to the strength of the pulsar
signal. When we artificially add gaussian-distributed noise at this
level to the observation, we find no sign of the pulsar's (2.6-Hz)
rotation frequency in 35-minute segments of the data\footnote{For
telescopes with a lower sensitivity than Arecibo (e.g. Parkes) then
even if PSR B0656+14 were quite a bit closer it would not reveal it's
periodicity in a similarly long observation.}  (left panel of
Fig. \ref{distribution}) yet the brightest pulse (right panel of
Fig. \ref{distribution}) is easily detected ($18\sigma$).  Only in the
spectrum of the whole 1.8-hour observation --- which is significantly
longer than the continuous observations of the Parkes Multibeam Survey
--- is the periodicity marginally detectable. This means that a
distant PSR B0656+14 could only be found as a RRAT in a survey using
Arecibo, unless the pointings were unusually long.

\section{Implications and discussion}

We have shown that PSR B0656+14 could have been identified as an RRAT,
had it been at the typical distance of the known RRATs. We have no way
of telling whether its capacity to produce intense bursts of emission
right across its profile is related to its age, period, inclination,
or even its immediate galactic environment, since this behavior has
been found in no other pulsar. The pulse-energy distribution is not a
power-law, but is better fitted by a lognormal distribution and such
distributions are thought to be common for pulsars
(e.g. \citealt{cjd04}).

In a study of 32 pulsars, \citet{rit76} found that PSR B0950+08 shows
the highest degree of pulse-to-pulse intensity
variation. Nevertheless, the brightest pulse found in an extensive
study of its field statistics by
\citet{cjd04} is approximately 5 {\Eav}. Vela does not show pulses
brighter than 10 {\Eav} and only 0.5\% of the pulses are brighter than
3 {\Eav} \citep{jvkb01}. For PSR B0656+14 4\% of the pulses are
brighter than 3 {\Eav}. Therefore the emission of PSR B0656+14 appears
to be extremely erratic compared with both normal pulsars and pulsars
with giant micropulses.

One could wonder if more known pulsars show a similar kind of sporadic
bright emission. To answer this question one should analyse the pulse
energy distribution of a sample of pulsars. A complication would be
that longer than typical observations are required to detect the
presence of a tail of strong pulses. In a large survey for subpulse
modulation by \citet{wes06}, this pulsar was the only pulsar that
showed clear evidence for this kind of sporadic emission.

Our identification of PSR B0656+14 with RRATs implies that at least
some RRATs could be sources which emit pulses continuously, but over
an extremely wide range of energies. This is in contrast to a picture
(predicted by \citealt{zgd06}) of infrequent powerful pulses with
otherwise no emission. Therefore, if it indeed turns out that PSR
B0656+14 (despite its relatively short period) is a true prototype for
an RRAT, we can expect future studies to demonstrate that
RRATs emit much weaker pulses among their occasional bright bursts. We
would also predict that their integrated profiles will be found to be
far broader than the widths of the individual bursts, and will need
many thousands of bursts to stabilize.

Hopefully, radio observations of RRATs will soon be able to test these
predictions. These, together with the detection of more RRATs and
potentially their high-energy counterparts, will shed light on their
true nature. The transient nature of these sources makes them
difficult to detect. However it is likely that the Galactic population
exceeds that of the ``normal'' radio pulsars \citep{mll+06}. Thus
surveys with long pointings, such as those planned with LOFAR, or many
observations of the same region of sky are required. Surveys at low
frequencies will also be more sensitive to nearer RRATs as the greater
degree of dispersion will allow them to be more easily distinguished
from radio frequency interference.

\acknowledgments

GAEW and JMR thank the Netherlands Foundation for Scientific Research
(NWO) and the Anton Pannekoek Institute, Amsterdam, for their kind
hospitality, and GAEW the University of Sussex for a Visiting
Fellowship. Portions of this work (for JMR) were carried out with
support from US National Science Foundation Grants AST 99-87654 and
00-98685. Arecibo Observatory is operated by Cornell University under
contract to the US National Science Foundation. The Westerbork
Synthesis Radio Telescope is operated by the ASTRON (Netherlands
Foundation for Research in Astronomy) with support from NWO.


\begin{thebibliography}{22}
\expandafter\ifx\csname natexlab\endcsname\relax\def\natexlab#1{#1}\fi

\bibitem[{Becker \& Tr\"{umper}(1997)}]{bt97}
Becker, W. \& Tr\"{umper}, J. 1997, A\&A, 326, 682

\bibitem[{{Brisken} {et~al.}(2003){Brisken}, {Thorsett}, {Golden}, \&
  {Goss}}]{btg+03}
{Brisken}, W.~F., {Thorsett}, S.~E., {Golden}, A., \& {Goss}, W.~M. 2003, ApJ,
  593, L89

\bibitem[{{Cairns}(2004)}]{cai04}
{Cairns}, I.~H. 2004, ApJ, 610, 948

\bibitem[{{Cairns} {et~al.}(2004){Cairns}, {Johnston}, \& {Das}}]{cjd04}
{Cairns}, I.~H., {Johnston}, S., \& {Das}, P. 2004, MNRAS, 353, 270

\bibitem[{{Cognard} {et~al.}(1996){Cognard}, {Shrauner}, {Taylor}, \& {Thorse
  tt}}]{cst+96}
{Cognard}, I., {Shrauner}, J.~A., {Taylor}, J.~H., \& {Thorse tt}, S.~E. 1996,
  ApJ, 457, L81

\bibitem[{{Haberl}(2004)}]{hab04}
{Haberl}, F. 2004, Advances in Space Research, 33, 638

\bibitem[{{Hankins} {et~al.}(2003){Hankins}, {Kern}, {Weatherall}, \&
  {Eilek}}]{hkwe03}
{Hankins}, T.~H., {Kern}, J.~S., {Weatherall}, J.~C., \& {Eilek}, J.~A. 2003,
  Nature, 422, 141

\bibitem[{Johnston \& Romani(2002)}]{jr02}
Johnston, S. \& Romani, R. 2002, MNRAS, 332, 109

\bibitem[{{Johnston} {et~al.}(2001){Johnston}, {van Straten}, {Kramer}, \&
  {Bailes}}]{jvkb01}
{Johnston}, S., {van Straten}, W., {Kramer}, M., \& {Bailes}, M. 2001, ApJ,
  549, L101

\bibitem[{{Knight} {et~al.}(2006){Knight}, {Bailes}, {Manchester}, {Ord}, \&
  {Jacoby}}]{kbmo+06}
{Knight}, H.~S., {Bailes}, M., {Manchester}, R.~N., {Ord}, S.~M., \& {Jacoby},
  B.~A. 2006, ApJ, 640, 941

\bibitem[{{Kramer} {et~al.}(2003){Kramer}, {Karastergiou}, {Gupta}, {Johnston},
  {Bhat}, \& {Lyne}}]{kkg+03}
{Kramer}, M., {Karastergiou}, A., {Gupta}, Y., {et~al.} 2003, A\&A, 407, 655

\bibitem[{{Kuzmin} \& {Ershov}(2004)}]{ke04}
{Kuzmin}, A.~D. \& {Ershov}, A.~A. 2004, A\&A, 427, 575

\bibitem[{Lorimer {et~al.}(1995)Lorimer, Yates, Lyne, \& Gould}]{lylg95}
Lorimer, D.~R., Yates, J.~A., Lyne, A.~G., \& Gould, D.~M. 1995, MNRAS, 273,
  411
\bibitem[{{McLaughlin} {et~al.}(2006){McLaughlin}, {Lyne}, {Lorimer}, {Kramer},
  {Faulkner}, {Manchester}, {Cordes}, {Camilo}, {Possenti}, {Stairs}, {Hobbs},
  {D'Amico}, {Burgay}, \& {O'Brien}}]{mll+06}
{McLaughlin}, M.~A., {Lyne}, A.~G., {Lorimer}, D.~R., {et~al.} 2006, Nature,
  439, 817

\bibitem[{{Popov} {et~al.}(2006){Popov}, {Turolla}, \& {Possenti}}]{ptp06}
{Popov}, S.~B., {Turolla}, R., \& {Possenti}, A. 2006, MNRAS, L34+

\bibitem[{{Reynolds} {et~al.}(2006){Reynolds}, {Borkowski}, {Gaensler}, {Rea},
  {McLaughlin}, {Possenti}, {Israel}, {Burgay}, {Camilo}, {Chatterjee},
  {Kramer}, {Lyne}, \& {Stairs}}]{rbg+06}
{Reynolds}, S.~P., {Borkowski}, K.~J., {Gaensler}, B.~M., {et~al.} 2006, ApJ,
  639, L71

\bibitem[{{Ritchings}(1976)}]{rit76}
{Ritchings}, R.~T. 1976, MNRAS, 176, 249

\bibitem[{{Shibanov} {et~al.}(2006){Shibanov}, {Zharikov}, {Komarova}, {Kawai},
  {Urata}, {Koptsevich}, {Sokolov}, {Shibata}, \& {Shibazaki}}]{szk+06}
{Shibanov}, Y.~A., {Zharikov}, S.~V., {Komarova}, V.~N., {et~al.} 2006, A\&A,
  448, 313

\bibitem[{{Soglasnov} {et~al.}(2004){Soglasnov}, {Popov}, {Bartel}, {Cannon},
  {Novikov}, {Kondratiev}, \& {Altunin}}]{spb+04}
{Soglasnov}, V.~A., {Popov}, M.~V., {Bartel}, N., {et~al.} 2004, ApJ, 616, 439

\bibitem[{{Weltevrede} {et~al.}(2006){Weltevrede}, {Edwards}, \&
  {Stappers}}]{wes06}
{Weltevrede}, P., {Edwards}, R.~T., \& {Stappers}, B.~W. 2006, A\&A, 445, 243

\bibitem[{{Woods} \& {Thompson}(2006)}]{wt06}
{Woods}, P.~M. \& {Thompson}, C. 2006, Compact Stellar X-ray Sources (Cambridge
  University Press), in press; astro-ph/0406133

\bibitem[{{Zhang} {et~al.}(2006){Zhang}, {Gil}, \& {Dyks}}]{zgd06}
{Zhang}, B., {Gil}, J., \& {Dyks}, J. 2006, ApJL, submitted ({\tt
  astro-ph/0601063})

\end{thebibliography}
\end{document}